\documentclass[aps,showpacs,nofootinbib,superscriptaddress,twocolumn,floatfix]{revtex4}

\usepackage{graphicx}
\usepackage{amssymb}
\usepackage{amsmath}
\usepackage{bm}
\usepackage{slashed}
\usepackage{datetime}
\usepackage{mciteplus}

\newcommand{\open}{\sphericalangle}

\begin{document}
\allowdisplaybreaks[2]

\title{
First glances at the transversity parton distribution through \\
dihadron fragmentation functions}

\author{Alessandro Bacchetta}
\email{alessandro.bacchetta@unipv.it}
\affiliation{Dipartimento di Fisica Nucleare e Teorica, Universit\`a di Pavia}
\affiliation{INFN Sezione di Pavia, via Bassi 6, I-27100 Pavia, Italy}

\author{Aurore Courtoy}
\email{aurore.courtoy@pv.infn.it}
\affiliation{INFN Sezione di Pavia, via Bassi 6, I-27100 Pavia, Italy}

\author{Marco Radici}
\email{marco.radici@pv.infn.it}
\affiliation{INFN Sezione di Pavia, via Bassi 6, I-27100 Pavia, Italy}

\begin{abstract}
We present first observations of the transversity parton distribution 
based on an analysis of pion-pair production in deep
inelastic scattering off transversely polarized targets. The extraction of
transversity relies on the knowledge of dihadron fragmentation functions,
which we take from electron-positron annihilation measurements. This is the
first attempt to determine the transversity distribution in the framework of
collinear factorization.
\end{abstract}



\maketitle

The distribution of quarks and gluons inside hadrons can be described by means of
parton distribution functions (PDFs). In a parton-model picture, PDFs describe
combinations of number densities of quarks and gluons in a fast-moving
hadron. The knowledge of PDFs is crucial for our understanding of Quantum
Chromodynamics (QCD) and for the interpretation of high-energy 
experiments involving hadrons. 

If parton transverse momentum is integrated over, in the Bjorken limit the partonic structure of 
the nucleon is described in terms of only three PDFs: the well known unpolarized, 
$f_1^q(x)$, and helicity, $g_1^q(x)$, distribution functions, and the transversity distribution 
function $h_1^q(x)$, which measures the transverse polarization of quarks with flavor $q$ 
and fractional momentum $x$ in a transversely polarized
nucleon~\cite{Ralston:1979ys,*Jaffe:1991kp,*Cortes:1991ja}.  
Intuitively, helicity and transversity give two orthogonal pictures of the partonic structure of 
polarized nucleons. They have very different properties, and transversity is much less known. 
In this work, we present an extraction of the $h_1^q$. 

Transversity is related to the interference of amplitudes with different helicities of partons and 
of the parent nucleon. In jargon, it is called a chiral-odd function. 
There is no transversity for gluons in a nucleon, and $h_1^q$ has a pure non-singlet 
scale evolution~\cite{Artru:1990zv}.  From transversity one can build the nucleon tensor charge, 
which is odd under charge conjugation and can be computed in lattice 
QCD~\cite{Gockeler:2006zu}  (for a review on transversity, see Ref.~\cite{Barone:2001sp} 
and references therein).
 
Transversity is particularly difficult to measure because it must 
appear in cross sections combined with another chiral-odd function. 
An
example is the cross section for single-particle inclusive Deep
Inelastic Scattering (DIS), where $h_1^q$ appears in a convolution with the
chiral-odd Collins fragmentation function  
$H_1^{\perp q}$~\cite{Collins:1993kk}, which describes the correlation between 
the transverse polarization of a fragmenting quark with flavor $q$ and the transverse momentum 
distribution of the detected unpolarized hadron. The convolution 
$h_1^q \otimes H_1^{\perp\, q}$ gives rise to a specific azimuthal
modulation of the cross section. The amplitude of the modulation has been
measured by the HERMES and COMPASS
collaborations~\cite{Airapetian:2010ds,*Alekseev:2010rw}. In order to extract
the transversity distribution from this signal, the Collins
function should be determined through the measurement of azimuthal asymmetries in 
the distribution of two almost back-to-back hadrons in $e^+e^-$ 
annihilation~\cite{Boer:1997mf}. The Belle collaboration has measured this 
asymmetry~\cite{Abe:2005zx,*Seidl:2008xc}, 
making the first-ever extraction of $h_1^q$ 
possible from a simultaneous analysis of $e p^{\uparrow} \to e' \pi X$ and 
$e^+ e^- \to \pi \pi X$ data~\cite{Anselmino:2007fs}. 

In spite of this achievement, some questions still hinder 
the extraction of transversity from single-particle-inclusive measurements. 
The most crucial issue is the treatment of evolution effects, since the measurements were 
performed at very different energies. The convolution  
$h_1^q \otimes H_1^{\perp\, q}$ involves the transverse momentum of quarks. 
Hence, its evolution should be described in the framework of the
transverse-momentum-dependent 
factorization~\cite{Collins:1981uk,*Ji:2004wu}. Quantitative explorations in this direction suggest
that neglecting evolution effects could lead to overestimating transversity~\cite{Boer:2008fr}. 

In this context, it is of paramount importance to extract transversity in an independent way, 
requiring only standard collinear factorization where the above complications are absent 
(see, e.g, Refs.~\cite{Collins:1989gx,*Brock:1993sz} and references therein). Here, we come 
for the first time to this result by considering the semi-inclusive deep-inelastic production of two 
hadrons with small invariant mass. 

In this case, the transversity distribution function is combined with a
chiral-odd Dihadron Fragmentation Function (DiFF), denoted as
$H_1^{\open\,q}$~\cite{Radici:2001na}, which describes the correlation between the 
transverse polarization of the fragmenting quark with flavor $q$ and the azimuthal 
orientation of the plane containing the momenta of the detected hadron pair. 
Contrary to the Collins mechanism, this effect survives after integration over quark transverse 
momenta and can be analyzed in the framework of collinear factorization.  This process has been 
studied from different perspectives in a number of
papers~\cite{Efremov:1992pe,*Collins:1994ax,Jaffe:1998hf,Radici:2001na,Bacchetta:2002ux}. 
The only published measurement of the relevant asymmetry has been presented by
the HERMES collaboration for the production of $\pi^+ \pi^-$ pairs on transversely polarized 
protons~\cite{Airapetian:2008sk}. Preliminary measurements have been presented
by the COMPASS collaboration~\cite{Wollny:2010}. Related preliminary
measurements in proton-proton scattering have been presented by the PHENIX
collaboration~\cite{Yang:2009zzr}.

Similarly to the single-hadron case, we need to independently determine $H_1^{\open\,q}$  
by looking at correlations between the azimuthal orientations of two pion pairs in back-to-back 
jets in $e^+e^-$ annihilation~\cite{Artru:1996zu,Boer:2003ya}. The measurement of this so-called 
Artru--Collins azimuthal asymmetry has recently  become possible thanks to the Belle 
collaboration~\cite{Vossen:2011fk}.   

In the HERMES publication~\cite{Airapetian:2008sk}, the asymmetry was denoted as 
$A_{UT}^{\sin(\phi_{R} + \phi_S)\sin \theta}$; for brevity and without ambiguity, here we will use the notation $A_{\text{DIS}}$. The data set was collected in bins of the variables $x$ (the momentum 
fraction of the initial quark), $z$ (the fractional energy carried by the $\pi^+ \pi^-$ pair), and 
$M_h$ (the invariant mass of the pair). Since our interest here lies mainly on the transversity 
distribution, and to avoid problems when dealing with three different projections of the same data 
set, we consider only the $x$ binning. In Tab.~\ref{t:HERMESdata}, we reproduce the data for 
convenience indicating for each bin also the average hard scale $\langle Q^2 \rangle$ and 
fractional beam energy loss $\langle y \rangle$.
\begin{table}
\begin{tabular}{c|c|c|c|c}
\hline\hline\\[-4.5mm]
bin boundaries & $\langle x \rangle$ & $\langle y \rangle$ & $\langle Q^2 \rangle$ (GeV$^2$)&
\( A_{\text{DIS}}\)  	\\[0.5mm]
\hline
0.023$<x<$0.040 & 0.033 & 0.734 & 1.232 & \( 0.015 \pm  0.010 \)  \\
0.040$<x<$0.055 & 0.047 & 0.659 & 1.604 & \( 0.002 \pm  0.011 \)  \\
0.055$<x<$0.085 & 0.068 & 0.630 & 2.214 & \( 0.035 \pm  0.011 \)  \\
0.085$<x<$0.400 & 0.133 & 0.592 & 4.031 & \( 0.020 \pm  0.010 \)  \\
\hline\hline
\end{tabular}
\caption{Semi-inclusive DIS data of the asymmetry \(  A_{\text{DIS}}\) from 
HERMES~\cite{Airapetian:2008sk}. The errors are mainly statistical (we added the systematic 
errors in quadrature). The average values of the variables are taken from Tab. 5.1 of 
Ref.~\cite{Lu:2008zzd}. The other variables have been integrated in the range
$0.5 \le M_h \le 1$ GeV and $0.2 \le z \le 1$.}
\label{t:HERMESdata}
\end{table}

The $A_{\text{DIS}}$ measured by HERMES~\cite{Airapetian:2008sk} can be interpreted 
as~\cite{Bacchetta:2006un}
\begin{equation} 
A_{\text{DIS}} (x, Q^2) =
- C_y\, 
\frac{\sum_q e_q^2\,h_1^q(x, Q^2)\, n_q^{\uparrow} (Q^2) }
        {\sum_q e_q^2\,f_1^q(x, Q^2) \, n_q (Q^2)} \; , 
\label{eq:asydis}
\end{equation} 
where (neglecting target-mass corrections)
\begin{equation} 
C_y =  \frac{\bigl\langle 1-y \bigr\rangle}
                    {\bigl\langle 1-y+y^2/2 \bigr\rangle}    \approx
\frac{1-\langle y \rangle}
       {1-\langle y \rangle +\langle y \rangle^2/2} \; .
\end{equation}
In Eq.~\eqref{eq:asydis}, we also introduced the following quantities
\begin{align} 
n_q (Q^2) &= \int  dz  \, dM_h^2\,D_1^{q\to\pi^+ \pi^-} (z, M_h^2, Q^2) \; ,  \nonumber
\\
n_q^{\uparrow} (Q^2) &= \int  dz \, dM_h^2\, 
\frac{|\bm{R}|}{M_h}\, H_{1, sp}^{\open q\to\pi^+ \pi^-} (z, M_h^2, Q^2) \; , 
\label{eq:npair}
\end{align}
with $|\bm{R}| / M_h =\sqrt{1/4 - m_\pi^2 / M_h^2}$ . 
$D_1^{q\to\pi^+ \pi^-}\!$ is the unpolarized DiFF describing the hadronization of a quark $q$ 
into a $\pi^+\pi^-$ pair plus any number of undetected hadrons, averaged over
quark polarization and pair orientation. Finally, $H_{1, sp}^{\open q\to\pi^+ \pi^-}\!$ is a chiral-odd 
DiFF, and denotes the component of $H_1^{\open q\to\pi^+ \pi^-}$ that is sensitive to the 
interference between the fragmentation amplitudes into pion pairs 
in relative $s$ wave and in relative $p$ wave, from which comes the common 
name of Interference Fragmentation Functions~\cite{Jaffe:1998hf}. 
Intuitively, if the fragmenting quark is moving along the $\hat{z}$ direction and is polarized 
along $\hat{y}$, a positive $H_{1, sp}^{\open q\to\pi^+ \pi^-}$ means that 
$\pi^+$ is preferentially emitted along $-\hat{x}$ and $\pi^-$ along $\hat{x}$.
Since in this case no ambiguities arise, in the following we shall conveniently simplify 
the notation by using $D_1^q$ and $H_1^{\open q}$ to denote the relevant DiFFs. 

In our analysis, we make the following assumptions (valid only for $\pi^+\pi^-$ pairs) based 
on isospin symmetry and charge conjugation~\cite{Bacchetta:2006un}:
\begin{gather}
D_{1}^{u} = D_{1}^{d} = D_{1}^{\bar{u}} = D_{1}^{\bar{d}} \; ,  \label{eq:ass1}
\\
D_1^s = D_1^{\bar{s}} \; , \quad D_1^c = D_1^{\bar{c}} \; , \label{eq:ass2}
\\
H_{1}^{\open u} = - H_{1}^{\open d} = - H_{1}^{\open \bar{u}} = 
H_{1}^{\open \bar{d}} \; , \label{eq:ass3}
\\
H_{1}^{\open s} = - H_{1}^{\open \bar{s}} = H_{1}^{\open c} = -
H_{1}^{\open \bar{c}} = 0 \; . \label{eq:ass4}
\end{gather}
We also assume $D_1^s  \equiv N_s\, D_1^u$ and we consider the two scenarios 
$N_s = 1$ and $N_s = 1/2$. The second choice is suggested 
by the output of the PYTHIA event generator~\cite{Sjostrand:2003wg}. 
Our final results will not depend strongly on this 
choice.  

The above assumptions allow us to turn Eq.~\eqref{eq:asydis} into the
following simple relation (neglecting charm quarks) 
\begin{equation} 
\begin{split} 
x h_1^{u_v}(x, Q^2) &- {\textstyle \frac{1}{4}}\, x h_1^{d_v}(x, Q^2) = 
- \frac{A_{\text{DIS}}(x, Q^2)}{C_y} \\
&\times  \frac{ n_u (Q^2) }{n_u^{\uparrow} (Q^2)} 
\sum_{q=u,d,s} \frac{e_q^2 N_q}{e_u^2} x f_1^{q+\bar{q}}(x, Q^2) \; , 
\end{split} 
\label{eq:simple}
\end{equation} 
where $N_u = N_d = 1$ and $f_1^{q+\bar{q}} = f_1^q + f_1^{\bar{q}}$,  
$h_1^{q_v} = h_1^q - h_1^{\bar{q}}$. 
Our goal is to derive from data the difference between the 
valence up and down transversity distributions by computing the r.h.s. of the above
relation. 
 
The PDFs in Eq.~(\ref{eq:simple}) can be estimated using any parametrization 
of the unpolarized distributions. We chose to employ the MSTW08LO PDF 
set~\cite{Martin:2009iq}. We checked that using different sets makes no
significant change. We also checked that the charm contribution is irrelevant.

The only other unknown term on the r.h.s.\  of Eq.~(\ref{eq:simple}) is the ratio 
$n_u / n_u^{\uparrow}$. We extract this information from the recent measurement 
by the Belle collaboration~\cite{Vossen:2011fk} of the Artru--Collins azimuthal asymmetry 
$A^{\cos (\phi_R + \bar{\phi}_R)}$~\cite{Artru:1996zu,Boer:2003ya,Bacchetta:2008wb} 
(denoted as $a_{12R}$ in the experimental paper). As in the previous case, without ambiguity 
we simplify the notation and refer to this asymmetry as $A_{e+e-}$. Using the 
assumptions~(\ref{eq:ass1})-(\ref{eq:ass4}), the asymmetry can be written as  
\begin{widetext}
\begin{equation} 
A_{e+e-} (z, M_h^2, \bar{z}, \bar{M}_h^2, Q^2) 
= - \frac{\langle \sin^2 \theta_2 \rangle}{\langle 1+\cos^2 \theta_2 \rangle} \, 
\frac{ \langle \sin \theta \rangle \, \langle \sin \bar{\theta} \rangle \, 5\, 
           \frac{|\bm{R}|}{M_h} \,H_{1}^{\open\, u} (z, M_h^2, Q^2)\,
            \frac{|\bar{\bm{R}}|}{\bar{M}_h} \, H_{1}^{\open\, u} (\bar{z}, \bar{M}_h^2, Q^2) }
        { (5+N_s^2)\, D_1^{u} (z, M_h^2, Q^2) \, D_1^u (\bar{z}, \bar{M}_h^2, Q^2)
           + 4\, D_1^c (z,M_h^2, Q^2) \, D_1^c (\bar{z}, \bar{M}_h^2, Q^2) }  \; ,
\label{eq:asye+e-}
\end{equation} 
\end{widetext}
where $\theta_2$ is the angle between the direction of the lepton annihilation and the 
thrust axis, and $\theta$ is the angle between the momentum of one 
hadron in the c.m. of the hadron pair and the total momentum of the pair in the 
laboratory~\cite{Bacchetta:2002ux}.

In Eq.~(\ref{eq:simple}) we need $n_u^{\uparrow} / n_u$ at the experimental values of 
$\langle Q^2 \rangle$ of Tab.~\ref{t:HERMESdata} and integrated over the HERMES 
invariant-mass range $0.5 \leq M_h \leq 1$ GeV. We will get to this number in two steps: first, 
we estimate the ratio $n_u^{\uparrow} / n_u$ integrated over $0.5 \leq M_h \leq 1$ GeV and 
at the Belle scale (100 GeV$^2$), then we address the problem of changing 
$Q^2$. 

We consider the Belle asymmetry integrated over $(z, \bar{z})$ and binned in 
$(M_h, \bar{M}_h)$. We restrict our attention only on the bins between 
$0.5 \leq (M_h, \bar{M}_h) \leq 1.1$ GeV (neglecting the small difference with the HERMES 
upper limit). We weight the contribution of each bin by the inverse of the statistical error squared, 
which should be to a good approximation proportional to the denominator of the asymmetry 
in each bin. By summing over all bins in the considered range, we get the total asymmetry 
\begin{equation}
\begin{split}
 A_{e+e-} &= \frac{-\langle \sin^2 \theta_2 \rangle}{\langle 1+\cos^2 \theta_2 \rangle} \, 
\frac{ \langle \sin \theta \rangle \, \langle \sin \bar{\theta} \rangle \, 5 \, (n_u^{\uparrow})^2}
        { (5+N_s^2)\, n_u^{2} + 4\, n_c^{2} }  \\
&= - 0.0307 \pm 0.0011 \; .
\label{eq:asye+e-int}
\end{split}
\end{equation} 
From the Belle analysis, we know that 
\begin{gather}
\frac{\langle \sin^2 \theta_2 \rangle}{\langle 1+\cos^2 \theta_2 \rangle} = 0.753 \; , \; 
\langle \sin \theta \rangle   \langle \sin \bar{\theta} \rangle = 0.871 \;  , \nonumber \\
\frac{4 n_c^2}{(5+N_s^2) n_u^2} = 0.415 \pm 0.047 \; . \nonumber
\end{gather} 
Therefore we obtain 
\begin{equation}
n_u^{\uparrow} / n_u (100\, \mathrm{GeV}^2) = - 0.273 \pm 0.007_{\mathrm{ex}} \pm 
0.009_{\mathrm{th}} \; , 
\label{eq:ratioQ0}
\end{equation} 
where the second error comes from using the two different values of the $s-$quark normalization 
$N_s$. We assumed the sign of the ratio to be negative in order to obtain a positive $u-$quark 
transversity distribution. To verify the reliability of this procedure, we repeated the calculation 
estimating the denominator of the asymmetry using the PYTHIA
event generator~\cite{Sjostrand:2003wg} 
without acceptance cuts (courtesy of the Belle collaboration). The result falls within the errors 
quoted in Eq.~(\ref{eq:ratioQ0}). 

The last step in the procedure is to address the $Q^2-$evolution of $n_u^{\uparrow} / n_u$, 
since the Belle scale is very different from the HERMES one (see Tab.~\ref{t:HERMESdata}). 
DiFFs must be connected from one scale to the other via their QCD evolution 
equations~\cite{Ceccopieri:2007ip}. In order to do this, we need to know the $z$ 
dependence of  $H_{1}^{\open\, u}$ and $D_1^q$ for each $M_h$ value. For $H_{1}^{\open\, u}$, 
we fit the Belle data for $A_{e+e-}$  binned in $(z, M_h)$ and integrated over 
$(\bar{z}, \bar{M}_h)$, multiplied by the inverse of the statistical error squared. 

The $D_1^q$ should be obtained from global fits of unpolarized cross sections, similarly to what is 
done for single-hadron fragmentation functions~\cite{deFlorian:2007aj}. In the absence 
of published data, we extract $D_1^q$ by fitting the unpolarized cross section as produced by 
the PYTHIA event generator~\cite{Sjostrand:2003wg}, 
which is known to give a good description of the total cross section.  
Following the assumptions introduced in Eqs.~(\ref{eq:ass1}) and (\ref{eq:ass2}), we describe 
the unpolarized cross-section for the production of a hadron pair with~\cite{Boer:2003ya}
\begin{equation}
\frac{d\sigma}{dz dM_h^2} = \frac{4 \pi \alpha^2}{Q^2} 
\left[ \frac{10}{9} D_1^{u} + \frac{2}{9} D_1^{s} + \frac{8}{9} D_1^c  \right] \; ,
\label{eq:ds0}
\end{equation}  
where $\alpha$ is the fine structure constant. We assume the integration over 
$\cos \theta_2$ to be complete in the Monte Carlo sample. 

We start from a parametrization of $D_1^q$ and $H_{1}^{\open\, u}$ at $Q_0^2 = 1$ GeV$^2$. 
Taking inspiration from the model analysis of Ref.~\cite{Bacchetta:2006un}, for both DiFFs 
we consider two channels which are effective in the considered range $0.5 \leq M_h \leq 1.1$ 
GeV: the fragmentation into the $\rho$ resonance decaying into $\pi^+\pi^-$,  and the 
continuum arising from the fragmentation into an incoherent $\pi^+\pi^-$ pair. Then, we evolve 
the DiFFs at LO using the HOPPET code~\cite{Salam:2008qg}, which we suitably extended to 
include also chiral-odd splitting functions. Finally, we fit the cross section~(\ref{eq:ds0}) and the 
numerator of the asymmetry~(\ref{eq:asye+e-}) in the bins of interest. We checked that the final 
results are affected in a negligible way by the gluonic component $D_1^g (z,M_h;Q_0^2)$. 
A thorough analysis will be presented in a future publication~\cite{Courtoy:2011}.  

By integrating the extracted DiFFs in the HERMES range $0.5 \le M_h \le 1$ GeV and 
$0.2 \le z \le 1$, we can calculate the evolution effects on $n_u^{\uparrow}/n_u$ at each 
$\langle Q^2 \rangle$ indicated in Tab.~\ref{t:HERMESdata}. It turns out that the ratio is 
decreased by a factor $0.92 \pm 0.08$, where the error takes into account the 
difference of $Q^2$ in the HERMES experimental bins as well as the uncertainty related to 
different starting parametrizations at $Q_0^2 = 1$ GeV$^2$. In conclusion, for the extraction of transversity in Eq.~(\ref{eq:simple}) we use 
the number
\begin{equation}
n_u^{\uparrow} / n_u = - 0.251 \pm 0.006_{\mathrm{ex}} \pm 0.023_{\mathrm{th}} \; .
\label{e:ratioQ}
\end{equation} 

\begin{figure}
\begin{center}
\includegraphics[width=0.5\textwidth]{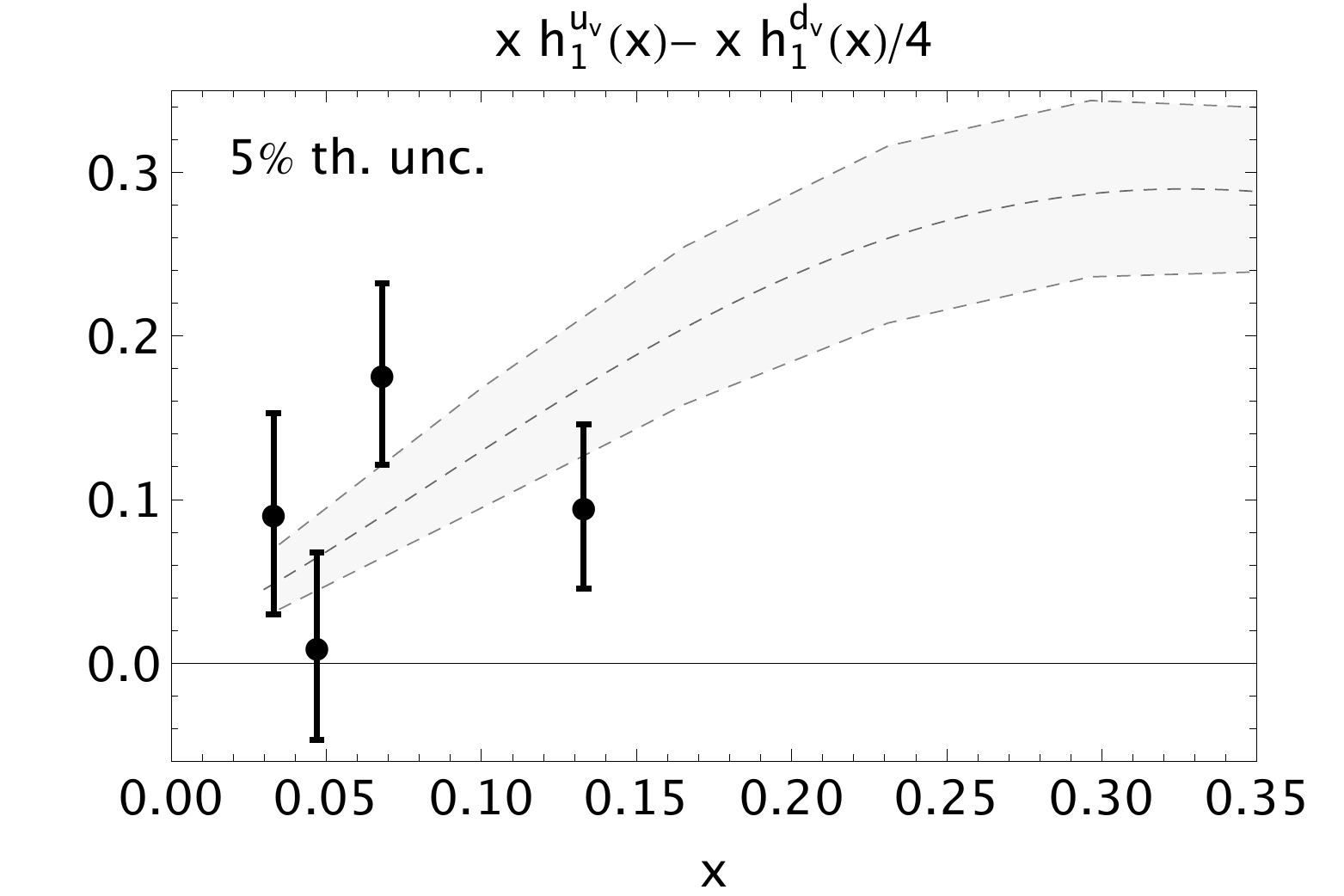}
\end{center}
\vspace{-0.5cm}
\caption{\label{fig:h1param} The $x h_1^{u_v} - x h_1^{d_v} / 4$ of Eq.~(\ref{eq:simple}) as a 
function of $x$. The error bars are obtained by propagating the statistical errors of each term in 
the equation. The uncertainty band represents the same observable as deduced from the 
parametrization of Ref.~\cite{Anselmino:2008jk}.}
\end{figure}
In Fig.~\ref{fig:h1param}, the data points denote the combination $x h_1^{u_v} - x h_1^{d_v} / 4$ 
of Eq.~(\ref{eq:simple}), plotted for each $\langle x \rangle$ and $\langle Q^2 \rangle$ listed in 
Tab.~\ref{t:HERMESdata}. We studied the influence of the errors of each element 
in the r.h.s.\  of Eq.~(\ref{eq:simple}). The only relevant contributions come from the 
experimental errors in the measurement of $A_{\text{DIS}}$, as reported in 
Tab.~\ref{t:HERMESdata}, and from the 9\% theoretical uncertainty on $n_u / n_u^{\uparrow}$. 

In Fig.~\ref{fig:h1param}, the central line represents the best fit for the combination 
$x h_1^{u_v} - x h_1^{d_v} / 4$, as  deduced from the most recent parametrization of 
$h_1^{u_v}$ and $h_1^{d_v}$ extracted from the Collins effect~\cite{Anselmino:2008jk}. 
The uncertainty band is obtained by considering the errors on the parametrization and 
taking the upper and lower limits for the combination of interest.  
Our data points seem not in disagreement with the extraction. However, a word of caution is 
needed here: while the error bars of our data points correspond to $1\sigma $ deviation from the 
central value, the uncertainty on the parametrization~\cite{Anselmino:2008jk} 
corresponds to a deviation $\Delta \chi^2 \approx 17$ 
from the best fit (see Ref.~\cite{Anselmino:2008sga} for 
more details). In any case, to draw clearer conclusions more data are needed (e.g., from the 
COMPASS collaboration~\cite{Wollny:2010}). 

In summary, we have presented for the first time a determination of the transversity parton 
distribution in the framework of collinear factorization by using data for pion-pair 
production in deep inelastic scattering off transversely polarized targets, combined with data 
of $e^+ e^-$ annihilations into pion pairs. The final trend of the extracted transversity 
seems not to be in disagreement with the transversity extracted from the Collins 
effect~\cite{Anselmino:2008jk}. More data are needed to clarify the issue. 

We thank the Belle collaboration and, in particular, Anselm Vossen for many fruitful 
discussions. We thank 
also Andrea Bianconi for many illuminating discussions. This work is partially supported by the 
Italian MIUR through the PRIN 2008EKLACK, and by the European Community through the 
Research Infrastructure Integrating Activity ``HadronPhysics2" (Grant Agreement n. 227431) 
under the $7^{th}$ Framework Programme. 


\bibliographystyle{apsrevM}
\bibliography{mybiblio}


\end{document}